\documentclass[showpacs,aps,amssymb,pra,amsmath]{revtex4}
\usepackage{float}
\usepackage{epsfig}
\usepackage{bm}
\usepackage{amsmath,graphicx}
\usepackage{subfig}
\usepackage{setspace}
\usepackage{color}

\newcommand{\beq}{\begin{equation}}
\newcommand{\eeq}{\end{equation}}
\newcommand{\bea}{\begin{eqnarray}}
\newcommand{\eea}{\end{eqnarray}}
\def\bk{{\bf k}}
\def\kv{{\bf k}}

\def\br{{\bf r}}
\def\rv{{\bf r}}

\begin{document}
\title{Testing the nonlocal kinetic energy functional of an inhomogeneous, two-dimensional degenerate Fermi gas within the average density approximation}

\author{J. Towers}
\affiliation{Jack Dodd Centre for Quantum Technology, Department of Physics, University of Otago, Dunedin, New Zealand}

\author{B. P. van Zyl}
\affiliation{Department of Physics, St. Francis Xavier University, Antigonish, Nova Scotia, Canada
B2G 2W5}
\author{W. Kirkby}
\affiliation{Department of Physics, St. Francis Xavier University, Antigonish, Nova Scotia, Canada
B2G 2W5}
%


\date{\today}
\begin{abstract}
In a recent paper [Phys.~Rev.~A {\bf 89}, 022503 (2014)], the average density approximation (ADA) was implemented to develop a parameter-free, nonlocal kinetic energy functional to be used in the orbital-free 
density-functional theory of an inhomogenous, two-dimensional (2D), Fermi gas.  In this work, we provide a detailed comparison of self-consistent calculations within the ADA  with the exact results
of the Kohn-Sham density-functional theory, and the elementary Thomas-Fermi (TF) approximation.  We demonstrate that the ADA for the 2D kinetic energy functional works very well under a wide variety of 
confinement potentials, 
even for relatively small particle numbers.  Remarkably,  the TF approximation for the kinetic energy functional, {\em without
any gradient corrections}, also yields good agreement with the exact kinetic energy for all confining potentials
considered, although at the expense of the spatial and kinetic energy densities exhibiting  poor point-wise agreement, particularly near the TF radius.
Our findings illustrate that the ADA kinetic energy functional yields accurate results for {\em both} the local and global equilibrium properties of an inhomogeneous 2D Fermi 
gas, without the need for any fitting parameters.  
\end{abstract}
\maketitle
\section{Introduction}
In a recent paper~\cite{towers_vanzyl}, the average density approximation (ADA) was used to construct a manifestly  nonlocal kinetic energy (KE) functional for a two-dimensional (2D), inhomogeneous
Fermi gas at zero-temperature.  The motivation behind this work was to provide an improved, orbital-free density-functional theory (DFT)~\cite{DFT}  beyond the 2D
Thomas-Fermi von Weizs\"acker (TFvW) theory, which 
relies upon an adjustable parameter, the von Weizs\"acker (vW) coefficient, $\lambda_{\rm vW}$,  for its implementation~\cite{PhD,zaremba_tso}.  
Owing to the fact that the TFvW  parameter needs to be fine-tuned depending on the
problem under consideration, the TFvW theory has the undesirable feature of not being universally applicable to an arbitrary inhomogeneous system.  
More importantly, in 2D, the TFvW theory is completely
{\em ad hoc}, with no formal justification within the usual gradient expansion techniques~\cite{kirkwood,kirzhnits57,jennings76,hodges73,geldhart86,salasnich,koivisto,PhD}.

While the ADA KE functional was shown in~\cite{towers_vanzyl} to yield superlative agreement with the exact results available for harmonic confinement, it is not immediately clear if similar agreement will be found for other
potentials~\cite{note1,brack_vanzyl}.  In this paper, we will examine the efficacy of the 2D KE functional under a variety of cofinement potentials, {\em via} fully self-consistent calculations for the noninteracting
system.  Our focus on the noninteracting system allows us to compare our ADA calculations with the results obtained within the Kohn-Sham (KS) DFT~\cite{DFT}, where the noninteracting KE is treated {\em exactly}.

To this end, the rest of our paper is organized as follows.  In the next section, we briefly review the ADA approach for constructing the 2D KE functional, and generalize our earlier results~\cite{towers_vanzyl} to also include the case of fully
spin-polarized Fermi gas.  In Sec.~\ref{numerics}, we present self-consistent  calculations under various confinement potentials, and compare the
spatially dependent  and integrated equilibrium properties with the exact KS DFT and the Thomas-Fermi (TF) approximation.  Our
central finding is that the ADA KE functional yields good agreement for both the local and global
equilibrium properties of a 2D inhomogeneous Fermi gas, with no adjustable parameters.
Finally, in Sec.~\ref{closing}, we present our closing remarks and directions for future work.
\section{The ADA and TFvW-like theory}\label{ADA}

The details of the derivation of the 2D nonlocal KE functional within the ADA have already been presented in Ref.~\cite{towers_vanzyl}.  Here, we will provide a simple generalization of this result, so that it may also be 
applicable to a spin-polarized
Fermi gas.  In order to take the spin degeneracy, $g$, into account ($g=1$ or $g=2$), we need only
introduce the following generalized quantities, all in 2D (hereby, we take $\hbar=m=1$ unless otherwise noted), {\it viz.,} the Fermi wave vector,
\beq\label{gen_kf}
|\bk_F| = k_F = \sqrt{\frac{4 \pi \rho}{g}}~,
\eeq
the KE per particle for the uniform system, 
\beq\label{gen_t0}
t_0 = \frac{\pi}{g} \rho~,
\eeq
and the static response function,
\beq\label{chi_0}
\chi_0(\eta) = -\frac{g}{2\pi}\begin{cases}
1,&\eta <1~,\\
{1-\sqrt{1-\dfrac{1}{\eta^2}}},&\eta \geq1~,
\end{cases}
\eeq
with 
\beq\label{gen_eta}
\eta \equiv \frac{|\bk|}{2k_F}~,
\eeq
and $\bk $ is  the wave vector conjugate to $\br -\br'$.
Following the  analysis carried out in Ref.~\cite{towers_vanzyl}
leads to the  2D nonlocal KE functional within the ADA, {\it viz.,} 
\beq\label{gen_Tnl}
T_{\rm ADA}[\rho] =  \frac{3\pi}{2g}\int d^2r \int d^2r'~\rho(\br')\tilde{w}(\br-\br';\rho(\br)){\rho}(\br) - \frac{1}{2}T_{\rm TF}[\rho] + T_{\rm vW}[\rho]~,
\eeq
where the vW KE functional is given by~\cite{vW}
\beq\label{TvW}
T_{\rm vW}[\rho] = \frac{1}{8}\int d^2r \frac{|\nabla \rho(\br)|^2}{\rho(\br)}~,
\eeq
and $T_{\rm TF}[\rho]$ is the TF KE functional, which reads

\beq\label{gen_TF}
T_{\rm TF}[\rho] = \frac{\pi}{g} \int d^2r \rho(\br)^2~.
\eeq
The quantity, $\tilde{w}(\br-\br';\rho(\br))$  in Eq.~\eqref{gen_Tnl} is the real-space nonlocal weight function~\cite{towers_vanzyl}, 
whose Fourier transform takes the form $\tilde{w}(\eta) \equiv \frac{2}{3} w_0 + \frac{1}{3}$, where
\begin{eqnarray}\label{weight2}
w_0(\eta) &=& \left[ {4\eta^2\ln\eta}+1+\left(\ln 4-3\right)\eta^2\right]\Theta(1-\eta) \nonumber\\ 
&+&\left[2\eta\sqrt{\eta^2-1} + \eta^2(\ln 4-2)  
-2\eta^2 \ln\left( 1+\sqrt{1-\frac{1}{\eta^2}}\right)\right]\Theta(\eta-1)~,
\end{eqnarray} 
and $\Theta(\cdot)$ is the Heaviside distribution.

The total energy functional is then given by 

\bea\label{gen_E}
E_{\rm ADA} [\rho] &=& \frac{3}{g}\int d^2r \int d^2r' \frac{\pi}{2} \rho(\br')\tilde{w}(\br-\br';\rho(\br))\rho(\br) 
-\frac{1}{g} \int d^2r \frac{\pi}{2}\rho(\br)^2 + \frac{1}{8} \int d^2r \frac{|\nabla \rho(\br)|^2}{\rho(\br)}\nonumber \\
&+& E_{\rm int}[\rho(\br)] +  \int d^2r~v_{\rm ext}(\br) \rho(\br)~,
\eea
where $v_{\rm ext}(\br)$ is some external potential, and $E_{\rm int}[\rho(\br)] $ encodes all of the interactions.  Note
that for $g=2$ (unpolarized), we obtain the results presented in Ref.~\cite{towers_vanzyl}.  Taking $g=1$ would be appropriate for investigating the
fully spin-polarized Fermi gas.

The variational minimization of Eq.~\eqref{gen_E} for a fixed number of particles yields the defining equations for what we have called the Thomas-Fermi von Weizs\"acker-like (TFvW-like) theory, {\it viz.,}
\beq\label{SC8}
-\frac{1}{2} \nabla^2\psi(\br) + v_{\rm eff}(\br) \psi(\br) = \mu \psi(\br)~,
\eeq
where we have introduced the vW wave function, $\psi(\br) \equiv \sqrt{\rho(\br)}$, and 
\beq\label{SC9}
v_{\rm eff}(\br) = -\frac{\pi}{g} \psi(\br)^2 + \phi(\br) + v_{\rm ext}(\br) + \frac{\delta E_{\rm int}}{\delta \rho(\br)}~,
\eeq
\bea\label{SC10}
\phi(\br)=\frac{3\pi}{2g} \int
\frac{d^2k}{(2\pi)^2}\int d^2 r_1
e^{i\kv\cdot(\rv-\rv_1)}\left[\Omega
\left(\frac{k}{2k_F(\rv)}\right)+ \tilde{w}
\left(\frac{k}{2k_F(\rv_1)}\right)\right] \rho(\rv_1)~,
\eea

\beq\label{NL3}
F(\eta) = \begin{cases}
1,&\eta <1~,\\
\dfrac{1}{1-\sqrt{1-\dfrac{1}{\eta^2}}},&\eta \geq1~,
\end{cases}
\eeq

\beq\label{Omega}
\Omega \left(\eta \right)\equiv \frac{2}{3} \left(F(\eta)+\frac{1}{2}-2 \eta^2\right)~.
\eeq
Note that we have made the local approximation, namely,  $k_F \to k_F(\br) = \sqrt{4\pi \rho(\br)/g}$ and $\eta \to \eta(\br) = k/2k_F(\br)$.
The normalization to the desired particle number, $N$, determines the chemical potential, $\mu$,
\beq\label{SC11}
N(\mu) = \int d^2r~|\psi(\br)|^2~.
\eeq
The numerical solution of the TFvW-like self-consistent equations is discussed in Ref.~\cite{towers_vanzyl}.  In what follows, we shall denote the self-consistent (SC) ADA equilibrium spatial density
by $\rho_{\rm sc}(\br)$.

\section{Comparison with exact Kohn-Sham calculations}\label{numerics}

By its definition, the KS kinetic energy
functional $T_{\rm KS}[\rho]$ of $N$-independent particles is (recall that $g=1,2$)
\beq\label{KSKE}
T_{\rm KS}[\rho] = g \sum_{i=1}^{N/g} \int d^2r~\phi_i^\star (\br)
\left(-\frac{1}{2}\nabla^2\right)\phi_i (\br)~.
\eeq
The $\phi_i(\br)$ are single-particle orbitals, each satisfying 
a Schr\"odingier-like equation 
{\it viz.,}
\beq\label{KSSE}
-\frac{1}{2}\nabla^2\phi_i(\br) + v_{\rm eff}(\br)\phi_i(\br) = \varepsilon_i\phi_i(\br)~,~~~~i=1,...,N
\eeq
where the effective potential is given by
\beq\label{KSVeff}
v_{\rm eff}(\br) \equiv  \frac{\delta E_{\rm int}[\rho]}{\delta\rho(\br)} + v_{\rm ext}(\br)~.
\eeq
The ground state density is obtained from
\beq\label{KSden}
\rho(\br) = g \sum_{i=1}^{N/g} \phi_i^\star(\br) \phi_i(\br)~,
\eeq
which at self-consistency, integrates to the total number of particles, $N$.  For a noninteracting system, 
the KS theory is {\em exact}.  Note that for $g=2$, we require an even number of fermions.  Furthermore, we also assume that the orbitals, $\phi_i(\br)$, are
fully occupied.  As  seen in Tables~\ref{table1}-\ref{table5} below, the requirement that the orbitals be fully occupied results in our KS calculations only being applicable to specific particle numbers,
depending on the confinement potential.

In what follows, we will take $g=2$, and compare the ADA equilibrium solutions for the noninteracting system with the
self-consistent  KS calculations, and the TF approximation .  We will examine the total energy,  the KE, as well as the quality of the equilibrium spatial and KE densities
({\it i.e.,} a point-wise comparison), in order
to determine how well the ADA performs under a wide variety of confinement potentials.  The inclusion of interactions
will not alter the general results found in this paper, since in either the TFvW-like, KS, or TF DFT, 
the same level of approximation is made for the interaction energy functional, $E_{\rm int}[\rho]$
({\it e.g.,} the Hartree-Fock approximation~\cite{DFT,vanzyl_pisarski,fang}).

\subsection{Power-law potentials}
We take as our  form for the power-law potential:
\beq\label{pot_power}
v_{\rm ext}(r) = \frac{1}{2} V_0 r^\alpha~,
\eeq
where $\alpha$ is a positive integer.
Here, we retain our fundamental constants ($\hbar$ and $m$), and rather opt to scale all lengths and energies by
$\ell = (\hbar^2/mV_0)^{1/(2+\alpha)}$ and $\hbar^2/m\ell^2$, respectively.  For example, with $\alpha=2$, and identifying
$V_0 = m \omega_0^2$, we obtain the usual harmonic oscillator (HO) length and energy scales.  For power-law potentials, Eq.~\eqref{pot_power}, the virial theorem yields,
\beq\label{power_virial}
T = \frac{\alpha}{2} V~,
\eeq
which results in the total energy, $E$, being given by $E = (1+ 2/\alpha)T$.  We may therefore focus our attention to a calculation of just the KE for power-law potentials.

In Tables~\ref{table1}-\ref{table4}, we compare the exact KS energies  with the ADA results, for $\alpha=2,4,6,8$, respectively, and several particle numbers. We have also included
an evaluation of $T_{\rm TF}[\rho_{\rm TF}]$, where $T_{\rm TF}[\rho]$ is given by Eq.~\eqref{gen_TF} and the TF energy functional is given by, 
\beq\label{E_TF}
E_{\rm TF}[\rho] = T_{\rm TF}[\rho] + \int d^2r~ v_{\rm ext}(\br) \rho(\br)~. 
\eeq
A variational minimization of Eq.~\eqref{E_TF} with respect to a fixed number of particles yields the TF density,
$\rho_{\rm TF}(r)$,  which in scaled units (hereby denoted by tildes) reads
\beq\label{rho_TF}
{\tilde \rho}_{\rm TF}(\tilde{r}) = \frac{1}{\pi}\left(\tilde{\mu}_{\rm TF} - \frac{1}{2}\tilde{r}^\alpha\right)
\Theta(\tilde{R}-\tilde{r})~,
\eeq
where $\tilde{R}$ is the TF radius and $\Theta(x)$ is the Heaviside distribution.  Note
that Eq.~\eqref{power_virial} is also obeyed in the TF approximation.

Table~\ref{table1} illustrates the agreement between the ADA, TF,  and KS 
calculations under harmonic confinement ($\alpha=2$).   It is clear that the nonlocal ADA KE functional performs quite well at reproducing the total energy of the
system, even for as few as $N=2$ particles.  Here, we have used the relative 
percentage error (RPE),
\beq\label{RPE}
{\rm RPE} \equiv \frac{|E_{\rm KS} - E|}{E_{\rm KS}}~,
\eeq
to quantify the agreement.  In Eq.~\eqref{RPE}, $E$ denotes the total energy calculated within the appropriate approximation ({\it i.e.,} the ADA or TF approximation).
It is evident from Table~\ref{table1}  that both the ADA and TF KE energies approach the KS KE energy as $N\to \infty$~\cite{towers_vanzyl}.
Note that for the isotropic HO in 2D, the particle numbers in Table~\ref{table1} correspond to the case of {\em closed shells}.  
It is also worthwhile pointing out just how well the TF approximation ({\it i.e.,} local-density approximation) does at giving the kinetic energy.  In fact, excluding $N=6$, the TF approximation yields
better results for the KE (compare the last two columns in Table~\ref{table1}) than the ADA, in spite of the fact that the TF spatial density, Eq.~\eqref{rho_TF}, possesses an unphysical sharp drop to zero
at the TF radius, ${\tilde R} = \sqrt{2}N^{1/4}$ with $\alpha=2$.

\begin{table}[ht] 
\centering      
\begin{tabular}{c c c c c c c c }  
\hline\hline                        
$N$ & $T_{\rm KS}$ & $T_{\rm ADA}[\rho_{\rm sc}]$ & $T_{\rm TF}[\rho_{\rm TF}]$ &  RPE$^{\rm ADA}$ & RPE$^{\rm TF}$  \\ [0.5ex] 
\hline                    
2    &  1   &  1.07  & 0.94 &7.0 & 6.0 \\
6    &  5   &  4.93  & 4.90  &1.4 & 2.0 \\
12    &  14   &   13.74    & 13.86 &1.9 & 1.0 \\
20   &  30   &  29.51  & 29.81  & 1.6 & 0.63 \\
30    &  55   &   54.28     & 54.77 &1.3 & 0.4 \\
90   &  285 &    283.16    & 284.61&  0.64 & 0.14 \\
132 &    506   &     503.60    & 505.52 &   0.47 & 0.09\\ 
182 &  819 &  816.11 & 818.44 &  0.35 & 0.07 \\ 
 420 & 2870 & 2866.48  & 2869.15 & 0.12 & 0.03 \\[1ex]       
\hline     
\end{tabular} 
\caption{Comparison of the $\alpha=2$ exact KS kinetic energy , $T_{\rm KS}$, 
with the kinetic energy obtained from $T_{\rm ADA}[\rho_{\rm sc}]$, 
and $T_{\rm TF}[\rho_{\rm TF}]$.  
The last two columns 
give the relative percentage error (RPE) between 
$T_{\rm KS}$ and $T_{\rm ADA}[\rho_{\rm sc}]$ and $T_{\rm KS}$ and $T_{\rm TF}[\rho_{\rm TF}]$, respectively.
Energies are measured in units of $\hbar^2/m\ell^2=\hbar\omega_0$ for the 2D isotropic harmonic oscillator.} 
\label{table1}  
\end{table} 

\begin{figure}[ht]
\centering \scalebox{0.6}
{\includegraphics{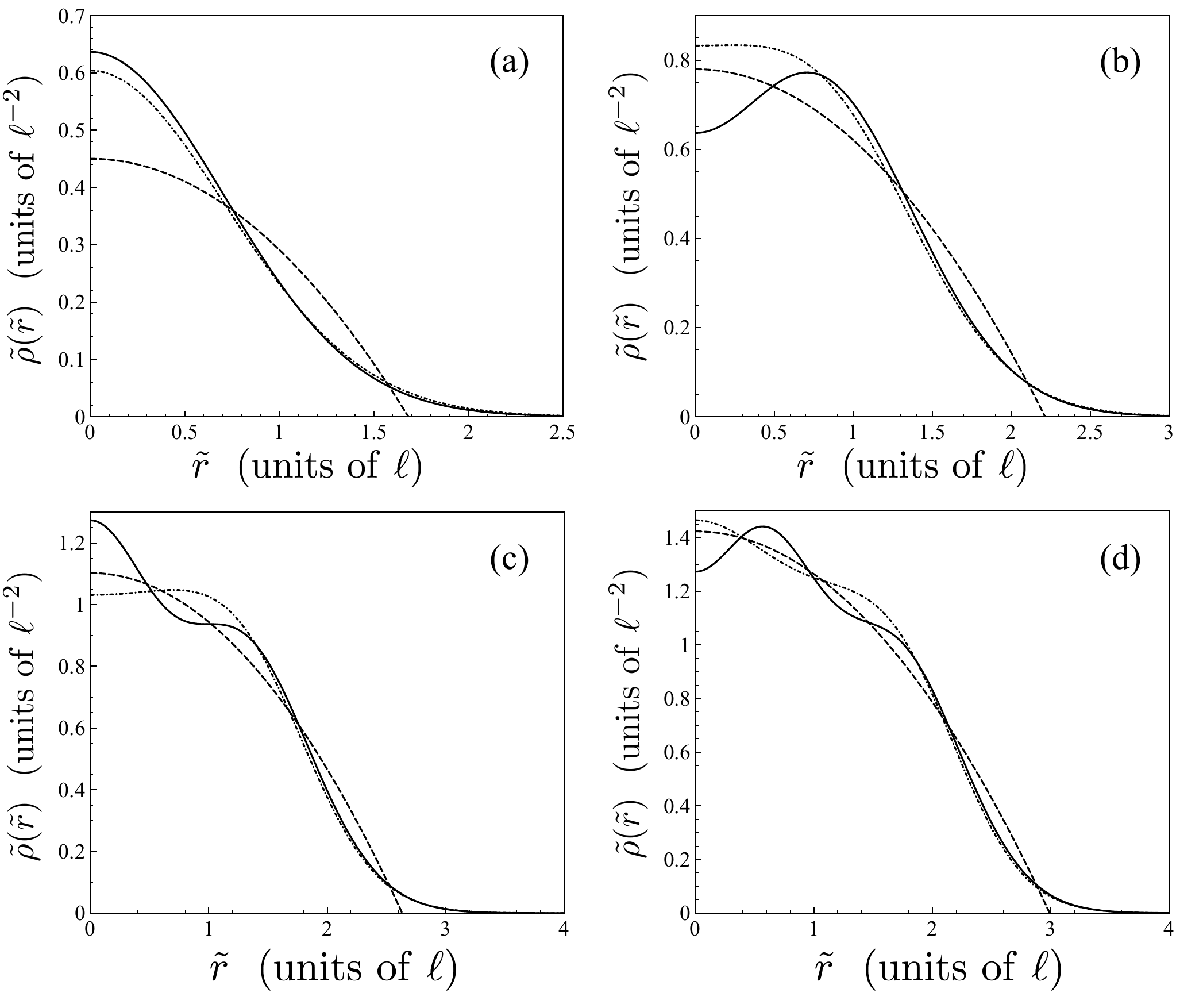}}
\caption{The spatial density, $\tilde{\rho}(\tilde{r})$ (scaled units) for the KS (solid curve), ADA (dot-dashed curve) and TF (dashed curve) evaluated at (a) $N=2$, (b) $N=6$, (c) $N=12$ and (d) $N=20$ particles with $\alpha=2$.  All 
quantities have been scaled as discussed in the text. }
\label{fig1}
\end{figure}

We can try to gain some insight into the nature of the two approximations by examining the spatial dependence of the
particle density for small particle numbers under harmonic confinement, illustrated in Fig.~\ref{fig1}.  
The most notable feature is how inadequate
the TF density (dashed curve)  is compared to the exact KS density (solid curve), especially in the low density region.  Indeed, the TF density is a monotonically
decreasing function of position regardless of the particle number, a property clearly not shared by the ADA densities for $N>6$.
Moreover,  the ADA (dot-dashed curve) provides excellent agreement with the KS density in the tail region.  In the
bulk, both the ADA and TF approximation fail to reproduce the oscillations in the KS KE density, although by
$N=20$, the ADA density shows signs of oscillatory structure, which while not directly attributable to the shell fillings,
is at least encouraging.  It is also interesting to note that both  the ADA and TF densities at the centre of the trap 
alternately lie below or above the KS density as one moves from panels (a)-(d).  The main message to be taken from
Fig.~\ref{fig1} is that good agreement for the energy (a global property) does not imply that the spatial density
will be likewise accurate; to wit, the TF approximation is far superior for obtaining the KE for $N>20$, but fails
completely to capture the surface profile of the quantum mechanical density, which is characterized by the low
density regime.

\begin{figure}[ht]
\centering \scalebox{0.6}
{\includegraphics{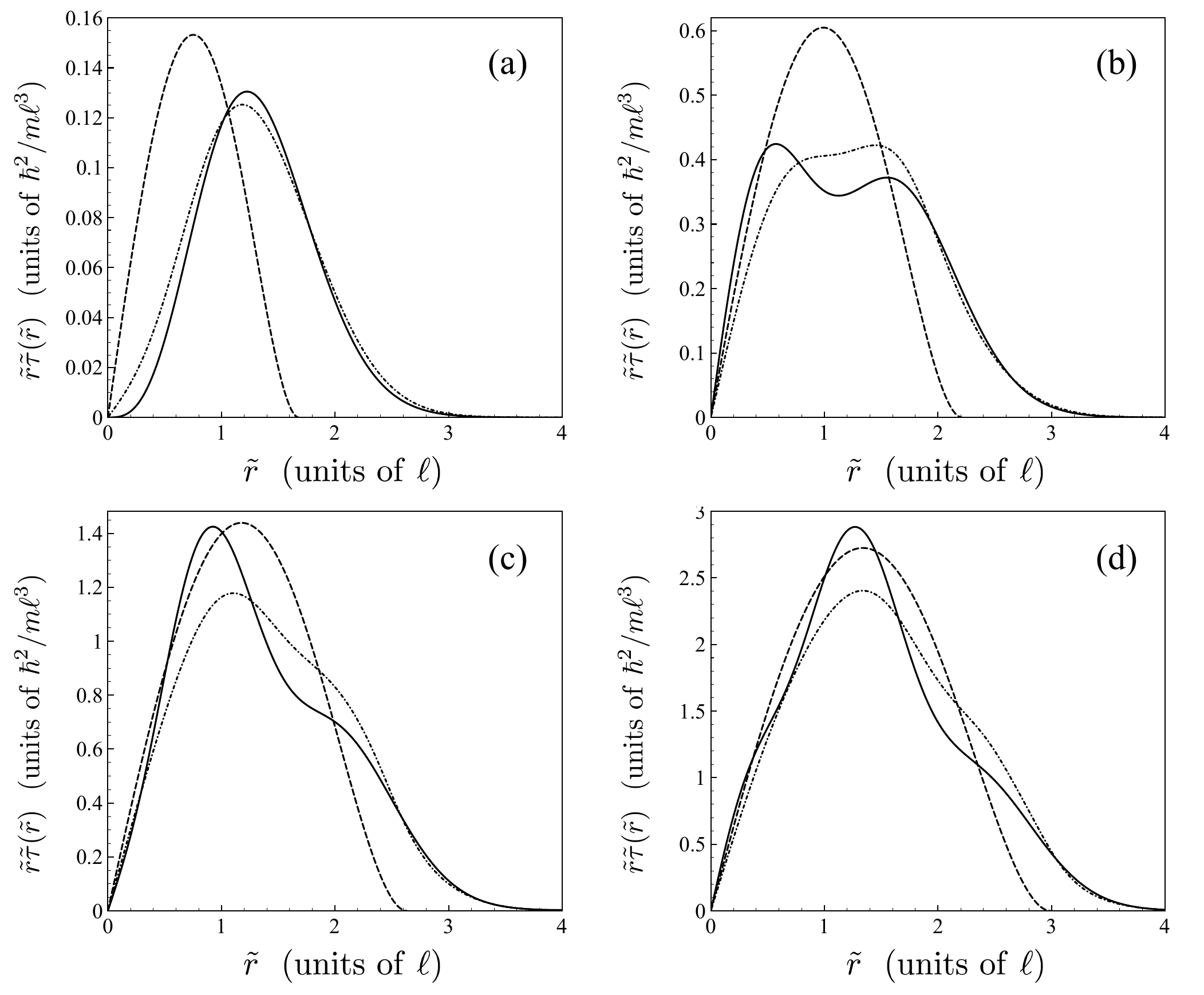}}
\caption{The kinetic energy density, $\tilde{r} \tilde{\tau}(\tilde{r})$ (scaled units) for the KS (solid curve), ADA (dot-dashed curve) and TF (dashed curve) evaluated at (a) $N=2$, (b) $N=6$, (c) $N=12$ and (d) $N=20$ particles with $\alpha=2$. }
\label{fig2}
\end{figure}

In Fig.~\ref{fig2}, we present the KE densities under harmonic confinement for the same number of particles
in Fig.~\ref{fig1}.  We have plotted $\tilde{r}\tilde{\tau}$ as this reflects the weighting of the radial variable in
the integral for calculating the KE.   It is immediately clear that that ADA provides a much better description of the
KE density than the TF approximation.  Indeed, for $N=2$ in panel (a), the TF KE density is remarkably poor, while
the ADA more closely follows the exact KS KE density, especially in the low-density region, where the ADA is noticeably superior to the TF approximation.  
As one moves up in particle number, the TF profile remains
qualitatively the same, while the ADA starts to show signs of the oscillatory structure present in the KS KE density.  Figure~\ref{fig2}
once again illustrates that simply examining the KE itself {\em is not} sufficient to determine the efficacy of any
proposed KE functional.  For example,  in panel (a), in spite of the TF KE density being noticeably inferior, the KE
it yields is actually {\em better} than the ADA KE.

\begin{table}[ht] 
\centering      
\begin{tabular}{c c c c c c c c }  
\hline\hline                        
$N$ & $T_{\rm KS}$ & $T_{\rm ADA}[\rho_{\rm sc}]$ & $T_{\rm TF}[\rho_{\rm TF}]$ &  RPE$^{\rm ADA}$ & RPE$^{\rm TF}$  \\ [0.5ex] 
\hline                    
2    &  1.56   &   1.70    & 1.32  & 9.0 & 15 \\
6    &  8.76   &   8.69    & 8.24  & 0.80 & 5.9 \\
10    &  20.66   &   19.63    & 19.31  & 5.0 & 6.5 \\
12    &  27.01  &   26.38    & 26.17  & 2.3 & 3.1\\
16    & 44.14  &  42.268    & 42.265  & 4.3 & 4.3 \\
20   & 62.55 &   61.06    & 61.30 & 2.4 &2.0\\
50 &    285.83   &   279.84    &282.31&  2.1 &1.2\\ 
60 &    385.76   &     379.29    &382.56 &  1.7& 0.83\\ 
300 &  5603.75 &  5570.70 & 5593.02 &  0.59 & 0.19\\ 
600 & 17773.39 & 17712.18 & 17756.70 &  0.34 & 0.09 \\
800 & 28699.42 & 28622.40 & 28681.00 & 0.27 & 0.06 \\[1ex]
\hline     
\end{tabular} 
\caption{As in Table~\ref{table1}, but for $\alpha=4$.} 
\label{table2}  
\end{table}

In Tables~\ref{table2}-\ref{table4}, we present the analogous data to Table~\ref{table1}, but now for $\alpha=4,6,8$, respectively.  Let us first examine Table~\ref{table2}, for which $\alpha=4$.  In contrast
to $\alpha=2$ in Table~\ref{table1}, the ADA is now much closer to the exact KS KE energy, for a similar number of particles.  However, as the particle number grows, the TF approximation
once again provides a better value for the exact KS KE, and by $N=300$, the TF approximation actually outperforms the ADA quite significantly.  

\begin{figure}[ht]
\centering \scalebox{0.6}
{\includegraphics{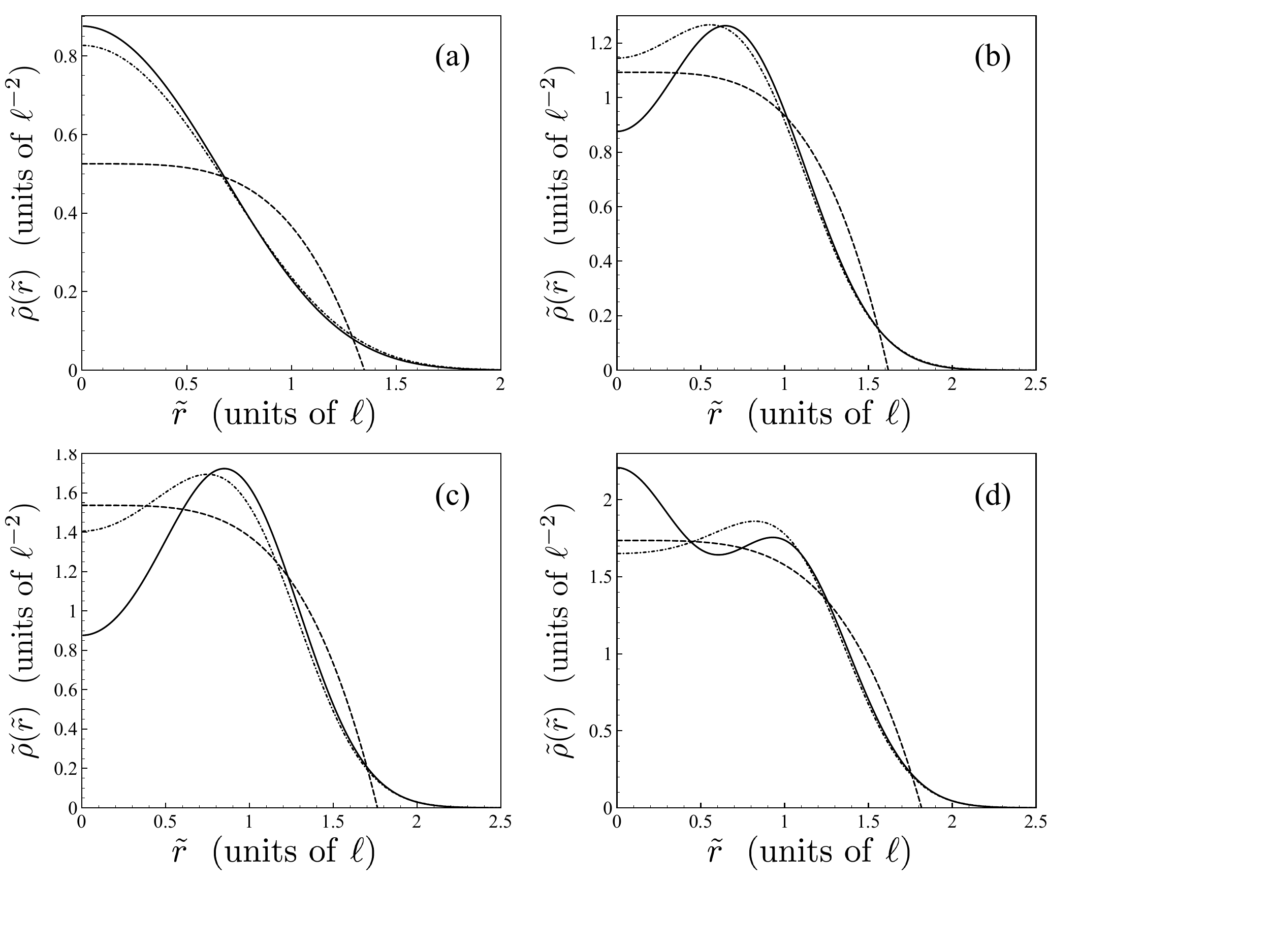}}
\caption{The spatial density, $\tilde{\rho}(\tilde{r})$ (scaled units) for the KS (solid curve), ADA (dot-dashed curve) and TF (dashed curve) evaluated at (a) $N=2$, (b) $N=6$, (c) $N=10$ and (d) $N=12$ particles with $\alpha=4$. All quantities
 are scaled as discussed in the text. }
\label{fig3}
\end{figure}

An examination of the spatial density distribution (Fig.~\ref{fig3})
reveals that even though the TF approximation yields reasonable agreement with the exact KS KE, it is an exceedingly poor representation of the exact spatial density (solid curves).  Indeed, for
$N=2$, the TF approximation (dashed curve) shares almost no features of the exact density, while the ADA (dot-dashed curve) does an admirable job, especially in the tail region.  
In addition, the TF densities in Fig.~\ref{fig3} appear to be much worse than for the HO confinement, which is perhaps expected given that the TF approximation for the density is only valid
for weakly inhomogeneous potentials.  In other words, as $\alpha$ is increased, the potential provides a very tight confinement, and as a result, the spatial TF density is unsurprisingly inferior.

\begin{figure}[ht]
\centering \scalebox{0.6}
{\includegraphics{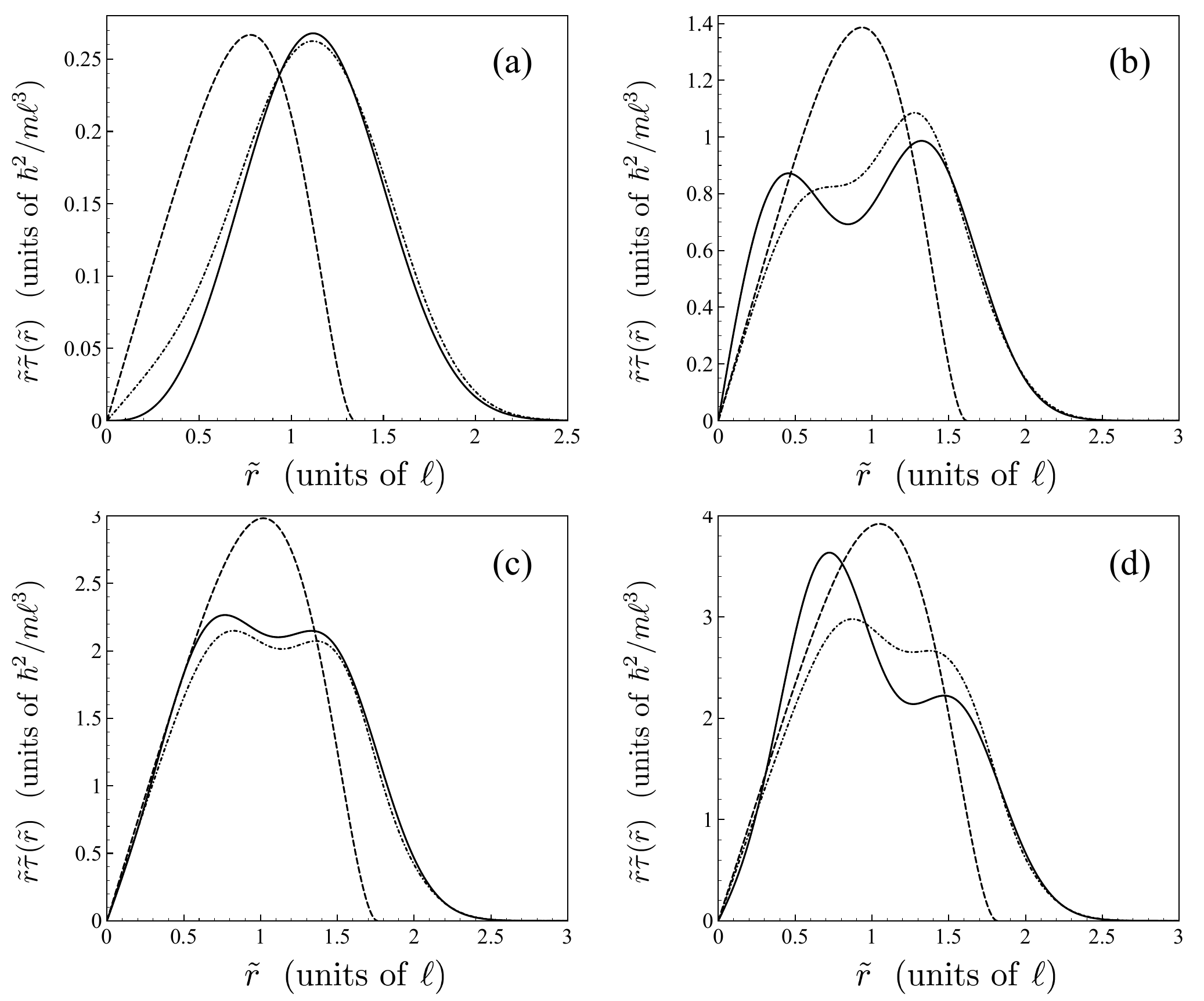}}
\caption{The kinetic energy density, $\tilde{r} \tilde{\tau}(\tilde{r})$ (scaled units) for the KS (solid curve), ADA (dot-dashed curve) and TF (dashed curve) evaluated at (a) $N=2$, (b) $N=6$, (c) $N=10$ and (d) $N=12$ particles with $\alpha=4$. }
\label{fig4}
\end{figure}

Similarly, in 
Fig.~\ref{fig4} we present the KE densities and observe the unsatisfactory behaviour of the TF approximation (dashed curves).  In contrast, the ADA KE density (dot-dashed curves) is in very good
agreement with the exact KS KE density (solid curves), including the oscillatory structure, particularly in panel (c), where the agreement is outstanding.  

\begin{table}[ht] 
\centering      
\begin{tabular}{c c c c c c c c }  
\hline\hline                        
$N$ & $T_{\rm KS}$ & $T_{\rm ADA}[\rho_{\rm sc}]$ & $T_{\rm TF}[\rho_{\rm TF}]$ &  RPE$^{\rm ADA}$ & RPE$^{\rm TF}$  \\ [0.5ex]  
\hline                    
2    &  1.96   &   2.15    & 1.50  & 9.7 & 23 \\
6    &  11.40   &  11.37    & 10.29 & 0.26 & 9.7 \\
10    &  27.64   &   26.30    & 25.15  & 4.9 & 9.0 \\
12    &  36.60   &   35.72   & 34.60  & 2.4 & 5.5 \\
16   &  60.67   &   58.18    & 57.24  & 4.1 & 5.7 \\
20   & 87.73 &    85.27    & 84.58 &  2.8 & 3.6 \\
50 &    427.98   &     417.74    & 420.40 &  2.4 & 1.8\\ 
60 &    585.55   &     574.35    &578.41  &   1.9 & 1.2 \\ 
300 &  9696.63 &  9624.13 &  9670.12 &   0.75 & 0.27\\ 
600 &  32572.81 &  32423.71 &  32526.3   & 0.46 & 0.14\\ 
700 &  42651.33 &  42476.65 &  42598.21  & 0.41 & 0.12\\[1ex]
\hline     
\end{tabular} 
\caption{As in Table~\ref{table1}, but for $\alpha=6$.} 
\label{table3}  
\end{table}

\begin{table}[ht] 
\centering      
\begin{tabular}{c c c c c c c c }  
\hline\hline                        
$N$ & $T_{\rm KS}$ & $T_{\rm ADA}[\rho_{\rm sc}]$ & $T_{\rm TF}[\rho_{\rm TF}]$  & RPE$^{\rm ADA}$ & RPE$^{\rm TF}$  \\ [0.5ex] 
\hline                    
2    &   2.26  &   2.50    & 1.61  & 11 & 29 \\
6    &  13.42   &   13.41    & 11.64  & 0.07 & 13 \\
10    &  32.93   &   31.38   & 29.18  &  4.7 & 11 \\
12    &  43.89   &   42.83    & 40.52  & 2.4 & 8 \\
16    &  73.27   &   70.36    & 68.01  & 4.0 & 7 \\
20   & 107.13 &    103.88    & 101.62 &  3.0 & 5.1 \\
50 &    541.80   &     527.64   & 528.79 &    2.6 & 2.4 \\ 
60 &    746.99   & 731.32       & 734.20 &   2.0 & 1.7 \\ 
300 &  13351.06 &  13236.05 & 13303.20 &  0.86 & 0.36 \\ 
700 &  61248.58 &   60940.60 & 61138.70 &  0.51 & 0.18 \\ 
900 &  96247.39 &  95844.85 & 96111.64 &  0.42 & 0.14 \\ 
\hline     
\end{tabular} 
\caption{As in Table~\ref{table1}, but for $\alpha=8$.} 
\label{table4}  
\end{table}

Tables~\ref{table3} and \ref{table4} correspond to $\alpha=6,8$, respectively, and follow the general trends observed in Table~\ref{table2}.  For this reason, we have not presented the spatial
densities and KE densities for these values of $\alpha$, and simply state that curves similar to those found in Figs.~\ref{fig1}-\ref{fig4} are found.  It is comforting to note that the ADA
continues to provide very good agreement in the tail region, even for large values of $\alpha$ ({\it i.e.,} steep confining potentials), indicating that the approximation is robust, and can 
provide accurate information about the properties of the system in the classically forbidden region ({\it e.g.,} this may be important in calculations of the binding energies, or atomic
polarizabilities, where the results are sensitive to the spatial density distribution).

\subsection{Shifted P\"oschl-Teller Potential}
For our final potential, we will examine the so-called 
shifted P\"oschl-Teller potential (PTP)\cite{poschl}, which can be written in the form
\beq\label{PT}
v_{\rm ext}(r) = V_0[1- {\rm sech}^2(r/a r_0) ]~,
\eeq
where $V_0>0$ is the depth, $r_0>0$ is the characteristic range of the potential, and $a >0$ is a dimensionless constant. 
Equation~\eqref{PT} differs fundamentally from the power law potentials studied above, since it has only a finite number of bound states owing to its saturation
to the constant value, $V_0$, as $r \to \infty$.   
The PTP also has uses in chemistry, where anharmonic potential energy surfaces are often needed for a realistic 
description of molecular dynamics, and in solid-state physics, where for example, the unshifted PTP,
{\it viz.,} $v_{\rm ext}(r) = -V_0 ~{\rm sech}^2(r/a r_0)$, is often used
to describe the ``hole'' produced by the diffusion of material species at double heterojunctions~\cite{kelly}.   

Scaling lengths by $r_0$, and energies by $\hbar^2/mr_0^2$ provides the dimensionless form for Eq.~\eqref{PT},
\beq\label{dimPT}
{\tilde v}_{\rm ext}(\tilde{r}) = {\tilde V}_0 [ 1 - {\rm sech}^2(\tilde{r}/a)]~.
\eeq
A series expansion of Eq.~\eqref{dimPT} about  $\tilde{r}/a$ contains even powers of $\tilde{r}/a$, implying that  the PTP can be viewed as a linear combination
of the even-$\alpha$ power-law potentials in Eq.~\eqref{pot_power}.
We choose the values $\tilde{V}_0=60$ and $a=6$ so that 
{\it (i)} the decay of the spatial density is on the same order as  the HO spatial density for similar particle numbers, and {\it (ii)} the potential can support
enough bound states to contain the desired range of particle numbers listed in Table~\ref{table5}.  


\begin{table}[ht] 
\centering      
\begin{tabular}{c c c c c c c c c c c}  
\hline\hline                        
$N$ & $T_{\rm KS}$ & $T_{\rm ADA}[\rho_{\rm sc}]$ & $T_{\rm TF}[\rho_{\rm TF}]$ & $E_{\rm KS}$ & $E_{\rm ADA}[\rho_{\rm sc}]$ & $E_{\rm TF}[\rho_{\rm TF}]$ & RPE$^{\rm ADA}$ & RPE$^{\rm TF}$  \\ [0.5ex] 
\hline                    
20   & 52.18 &    51.38    & 51.96 & 106.95 & 105.24 & 106.39 & 1.6 & 0.52 \\
60 &    261.51   &     258.75 & 260.49 &   546.34 & 539.63 & 543.38  & 1.2 & 0.54 \\ 
120 &  711.13 & 707.61 & 710.34 &  1514.57 & 1504.09& 1510.61  & 0.69 & 0.26\\ 
240 &  1904.12 &  1899.00 & 1902.57 &  4169.03 & 4156.70& 4166.98  & 0.30 & 0.05\\ 
520 &  5511.52 &  5509.30 & 5512.31 &  12745.38 & 12726.94& 12742.00 & 0.14  & 0.03 \\ 
670 &  7713.45 &  7712.60 & 7714.49 &  18297.86 & 18279.01& 18295.41  & 0.10  & 0.01 \\ [1ex]
\hline     
\end{tabular} 
\caption{Comparison of the exact KE,  $T_{\rm KS}$, and the exact KS total energy, $E_{\rm KS}$, 
with the same quantities calculated {\it via} the ADA and TF calculations, under the PTP confinement, Eq.~\eqref{PT}.  
The last two columns 
give the relative percentage error (RPE) between 
$E_{\rm KS}$ and $E_{\rm ADA}[\rho_{\rm sc}]$ and $E_{\rm KS}$ and $E_{\rm TF}[\rho_{\rm TF}]$, respectively.
Energies are measured in units of $\hbar^2/mr_0^2$.} 
\label{table5}  
\end{table} 

\begin{figure}[ht]
\centering \scalebox{0.6}
{\includegraphics{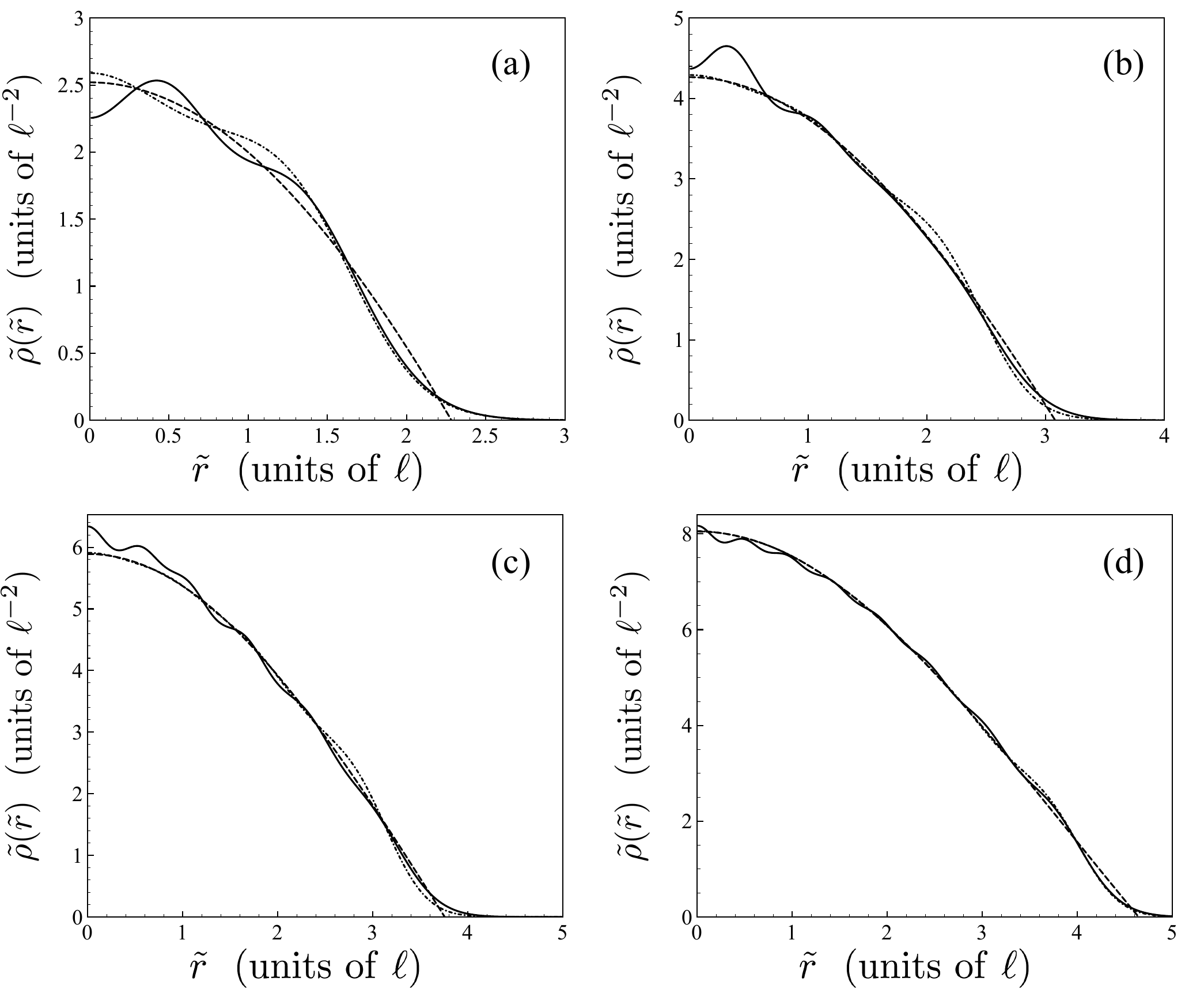}}
\caption{The spatial density, $\tilde{\rho}(\tilde{r})$ (scaled units) for the KS (solid curve), ADA (dot-dashed curve) and TF (dashed curve) evaluated at (a) $N=20$, (b) $N=60$, (c) $N=120$ and (d) $N=240$ particles for the shifted PTP. }
\label{fig5}
\end{figure}

\begin{figure}[H]
\centering \scalebox{0.6}
{\includegraphics{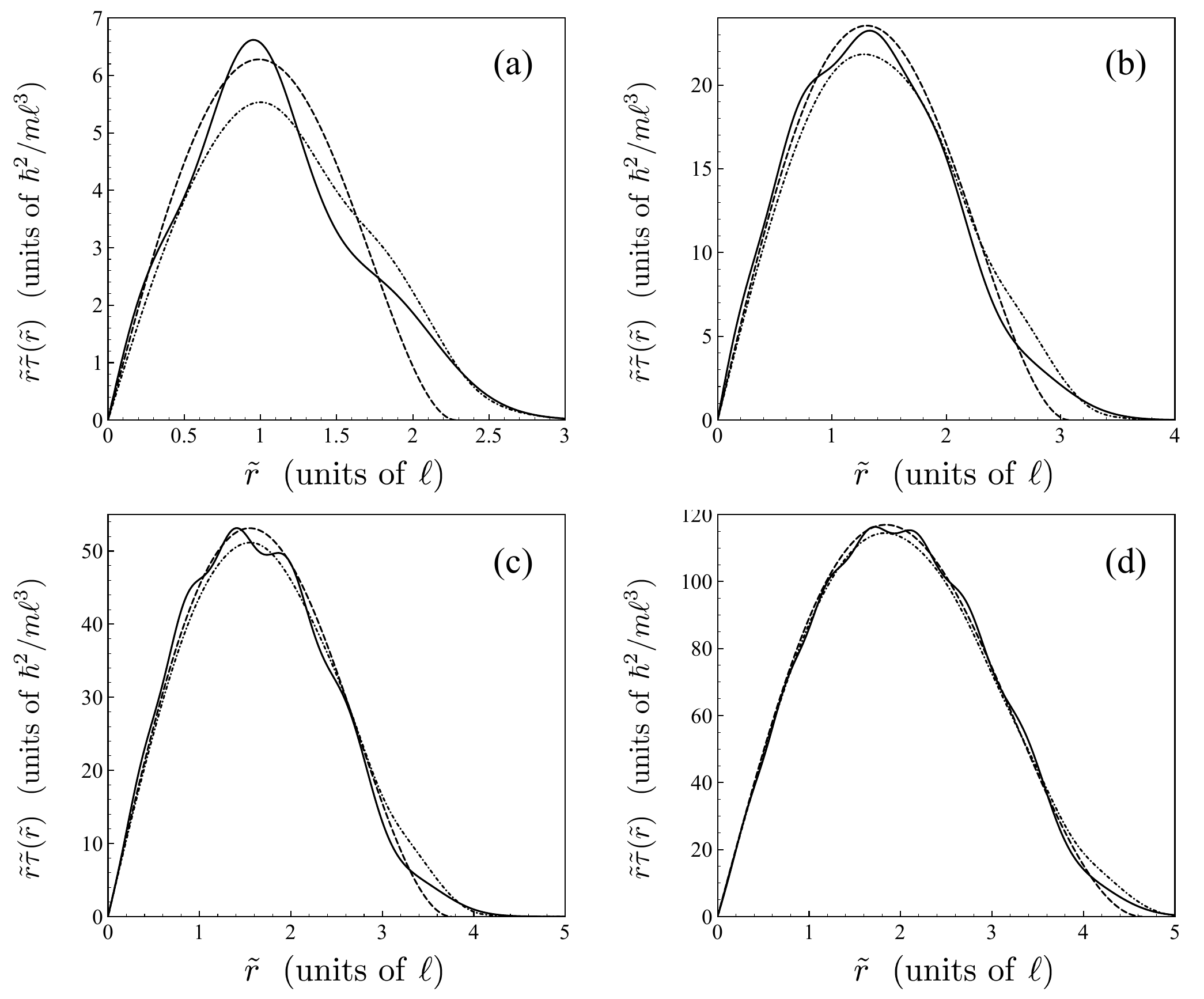}}
\caption{The kinetic energy density, $\tilde{r} \tilde{\tau}(\tilde{r})$ (scaled units) for the KS (solid curve), ADA (dot-dashed curve) and TF (dashed curve) evaluated at (a) $N=20$, (b) $N=60$, (c) $N=120$ and (d) $N=240$ particles for the shifted PTP. }
\label{fig6}
\end{figure}

Table~\ref{table5} provides a comparison between the exact KS energy, $E_{\rm KS}$, and KE, $T_{\rm KS}$,  with the same quantities calculated {\it via} the ADA  and TF approximation.
Again, we observe the trend that the TF approximation yields very good agreement with the exact KS energy, and as we increase the number of particles, the TF approximation improves.  By $N=670$
particles, the RPE for the TF approximation is already better by an order of magnitude.  Nevertheless, the TF approximation provides a comparatively inferior description of the spatial and KE densities as illustrated in Figs.~\ref{fig5} and \ref{fig6}; in contrast,
the ADA continues to give an accurate description of the densities in the very low density regime.  
It is also apparent that the ADA spatial densities closely follow the TF approximation in the bulk, with the important 
difference of having a smooth decay into the classically forbidden region, in good agreement with the exact KS
spatial density.

In Fig.~\ref{fig6}, we observe that even though the spatial densities associated with the TF and ADA are quite similar (panels
(b)-(d) in Fig.~\ref{fig5}),
the KE densities reveal some notable differences both within the bulk and the tail region. In particular, the ADA
KE density does not capture the oscillatory behaviour of the KS KE density as well as in the power law potentials, which 
highlights the fact that the oscillations in the ADA are cannot be directly associated with the filling of the orbitals in the
KS calculation.
\section{Closing Remarks}\label{closing}

We have provided a detailed examination of the efficacy of the recently proposed ADA for the KE functional of an inhomogeneous 2D Fermi gas.  Our central finding is that while even the exceedingly
simple TF KE functional can yield very good agreement with the exact KE of the system, it does a very poor job at providing the spatial and KE densities, especially at low particle numbers.  Our findings
also nicely illustrate that the 2D HO potential is a rather special case in 2D~\cite{note1} (see also the discussion in Ref.~\cite{towers_vanzyl}), and should not be the sole benchmark for testing the quality of any proposed KE functional for an inhomogeneous system.
On the other hand, the ADA KE functional we
have proposed for the 2D Fermi gas yields comparatively excellent agreement with exact KS calculations for the KE, as well as the local ({\it i.e.,} point-wise) densities, for a variety of trapping potentials.  Our results demonstrate
that the ADA provides an accurate, {\em parameter-free} KE functional which can be used with confidence in orbital-free DFT, even when the external potentials are tightly confining. 

Finally, it would be of interest to implement the work of Kugler~\cite{kugler} to go beyond the linear-response condition leading to our ADA KE functional, and utilize the hierarchy of coupled equations presented in 
Ref.~\cite{kugler}, which in principle fully
determine the KE functional for an arbitrary inhomogeneous system.  It is hoped that by including a truncated set of functional expressions, a closed system of equations can be solved, thereby developing
an improved KE functional, and providing insight into constructing a truly universal expression for the KE functional, which can be implemented in an orbital-free density-functional theory.

\acknowledgements
This work was supported by grants from the Natural Sciences and Engineering Research Council of Canada (NSERC).  W. Kirkby would like to thank the NSERC Undergraduate Summer Research Award
(USRA) for additional financial support.  B. P. van Zyl would like to thank E. Zaremba for useful discussions during the early stages
of this work.

%

\end{document}